\documentclass[lnbip]{svmultln}
\usepackage{makeidx}  
\usepackage{graphicx}
\usepackage{amsmath}
\usepackage{float}
\usepackage{diagbox}
\usepackage{tabularx}
\begin{document}
\mainmatter              

\title{RGDA-DDI: Residual graph attention network and dual-attention based framework for drug-drug interaction prediction}
\titlerunning{RGDA-DDI}  
%
\newcommand*\samethanks[1][\value{footnote}]{\footnotemark[#1]}
\author{Changjian Zhou\inst{1,2} \and Xin Zhang\inst{2,3} \and Jiafeng Li\inst{4}\and Jia Song\inst{1}\thanks{Corresponding author.} \and Wensheng Xiang\inst{1}\samethanks}
\authorrunning{Changjian Zhou et al.}   
%
\tocauthor{}
\institute{Key Laboratory of Agricultural Microbiology of Heilongjiang Province, Northeast Agricultural University, Harbin, China\\
\and
Dept. of Data and Computing, Northeast Agricultural University, Harbin, China\\
\and
School of Electrical and Information, Northeast Agricultural University, Harbin, China\\
\and
Galileo Financial Technologies, Utah, USA}

\maketitle              

\begin{abstract}        
Recent studies suggest that drug-drug interaction (DDI) prediction via computational approaches has significant importance for understanding the functions and co-prescriptions of multiple drugs. However, the existing silico DDI prediction methods either ignore the potential interactions among drug-drug pairs (DDPs), or fail to explicitly model and fuse the multi-scale drug feature representations for better prediction. In this study, we propose RGDA-DDI, a residual graph attention network (residual-GAT) and dual-attention based framework for drug-drug interaction prediction. A residual-GAT module is introduced to simultaneously learn multi-scale feature representations from drugs and DDPs. In addition, a dual-attention based feature fusion block is constructed to learn local joint interaction representations. A series of evaluation metrics demonstrate that the RGDA-DDI significantly improved DDI prediction performance on two public benchmark datasets, which provides a new insight into drug development.
\end{abstract}
\section{Introduction}
Drug-drug interaction (DDI) prediction plays an important role in modern biomedical and pharmaceutical research, as it is common for a patient to take multiple drugs at the same time, which provides therapeutic benefits but at the same time increases the potential risk of drug combination incompatibility [1]. Precision prediction of DDIs is still a very challenging task because of the large number of drug-drug pairs (DDPs). The traditional hand-crafted in vitro biomedical experiments such as safety profiling and clinical trials are reliable but time-consuming and labor-intensive, preventing their application on large-scale data [2]. To narrow down the search scope of compound candidates, in the past decade, several computational silico approaches have been proposed to identify high-confidence DDPs, and provide new insights into the causes of potential adverse side effects in drug combinations [3]. Most existing computational DDI prediction methods mainly focus on exploiting information from similarity-based features, or from homogeneous networks.\\

Similarity-based approaches assume that similar drugs may have interactions with each other, and these approaches require hand-extracted features being passed into the conventional machine learning based models, such as support vector machine (SVM) and random forest (RF), etc. For example, Vilar et al. [4] proposed a structural similarity based DDI prediction approach, and described a protocol with applications in patient safety and preclinical toxicity screening. Deng et al. [5] presented a molecular similarity-based machine learning framework to predict DDIs using various classification algorithms such as naive bayes, decision tree, random forest, and so on. Cheng et al. [6] extracted drug features such as phenotypic, chemical, therapeutic, and genomic properties for classification, and applied five predictive models in the framework: naive bayes, decision tree, k-means, logistic regression, and support vector machine, respectively. The results indicated that integrating multiple classifiers is feasible. \\

Hand-engineered features, however, require expert knowledge and intensive effort to extract the fine-grained features from omics data, and hence prevent the prediction architectures from being scaled to large-scale drug datasets. Indeed, the similarity-based models could add more complexity to the task at the data generation and processing stage already, involving processes such as digital imaging to extract features[7]. The network-based methods have surpassed these traditional models with their ability to capture useful latent features automatically, leading to highly flexible DDI prediction models. Ryu et al. [8] proposed a deep learning-based framework that uses drug structure information with principal components analysis (PCA) to reduce feature dimensions, then the concatenated feature vectors of drug pairs are fed into the deep neural network (DNN) to predict the type of DDIs. Huang et al. [9] applied a SVM framework to predict DDIs based on a long short-term memory (LSTM) model. The experimental results show that the proposed method has achieved a decent performance. In addition, transformer models also emerged and showed excellent performance in drug research related prediction task, for example, the TransVAE-DTA model developed by Zhou et al. [10] demonstrated superiority in drug-target binding affinity prediction. It is worth noting that, transformer models showed superior performance in general prediction tasks not only limited to drug research, but also, for example, agricultural disease identification[11]. Deng et al. [12] pointed out that, most models fail to reveal the DDI-associated events regardless of predicting the interaction or not. Consequently, they proposed a multimodal deep learning framework that combines diverse drug features with deep learning to build a model for predicting DDI-associated events.\\

In addition, the knowledge graph is widely used in DDI predictions, which can provide more detailed and comprehensive drug lateral side information. Ren et al. [13] presented a deep learning framework named DeepLGF to fully exploit biomedical knowledge graph (BKG) fusing local-global information to improve the performance of DDIs prediction. Md et al. [14] propose a novel DDIs prediction framework using knowledge graphs (KGs). This novel approach embeds the nodes in the graph using various embedding approaches. However, the knowledge graphs inevitably suffer from data noise pitfalls, which limits the performance and interpretability of models based on the knowledge graphs. To mitigate these limitations, Hong et al. [15] presented a link-aware graph attention approach for DDI prediction.  For a drug pair link, the novel method uses embedding representation as a query vector to calculate attention weights, and selects the appropriate topological neighbor nodes to obtain the semantic information of the other drugs. Together, these works greatly advanced our understanding of the prediction of DDIs in different species under various conditions. However, existing approaches suffer from the following limitations.\\

First, most existing studies only focus on a single drug molecular feature representation method, but fail to support multiple feature learning simultaneously through an integrated predictive model. Therefore, the study of the interplay between different drug interaction features is limited. Furthermore, most of the work in the field, only concatenate the two drug features for DDI prediction, but ignore the DDP features that contributed to the predictions.\\

With the reasoning above, there is a great motivation to take advantage of state-of-the-art deep learning techniques to develop a unified predictive framework that supports multiple feature learning simultaneously by integrating multiple feature fusion technologies. Here, we present RGDA-DDI, a residual-GAT and dual-attention based approach for integrated prediction and interpretation of DDIs. We use residual-GAT for this research as residual network based models has shown great performance in different prediction tasks, for example, a restructured deep residual dense network achieved average prediction accuracy of 95\% in the task of tomato leaf disease identification [16]. The architecture of our approach enables accommodation of the shared structure features of different drugs, and employs a dual-attention mechanism to gain insights of the trained model. A series of evaluation metrics were performed on two public benchmark datasets. Experimental results show that the RGDA-DDI method achieves better performance than the existing state-of-the-art baseline models, demonstrating that effective fusion of topological information and attribute information is beneficial. The main contributions of this study are summarized as follows.\\

• A novel residual graph attention network and dual-attention based framework (RGDA-DDI) is proposed for improving DDI prediction performance on a series of evaluation metrics. RGDA-DDI enables to train allows training the input on two drugs and DDPs simultaneously, make it take thus taking full advantage of drug and DDP features.\\

•  novel drug feature representation module is designed by utilizing residual concatenation between GAT and self-attention graph pooling (SAGPooling) to improve feature learning ability.\\

•  novel dual-attention module is presented for learning local interactions between the two encoded drug representations, which provides richer joint interaction representations while keeping the computational cost at the same scale.\\

\section{Related Work}
This study presents a novel residual-GAT and dual-attention based approach for DDI prediction, which aims to achieve a better performance than existing baseline models. The proposed method is closely related to two base models, graph attention network and biomedical knowledge graph (BKG). A brief review of these models is given as follows.\\

\subsection{Graph attention network}

The graph attention networks (GAT) [17] combines attention mechanisms [18] and graph convolutional networks (GCN) [19], which utilize an attention-based aggregator to generate attention weights over all neighbors of a node for feature learning. Like GCN, the aggregator function of GATs can be formed as Eq.(1).\\
\begin{equation}
\beta^{l+1}_{i}=\lambda(\sum_{j \in N_{i}} \alpha^{l}_{ij} \cdot\beta^{l}_{j}\cdot \omega^{l} )
\end{equation}
where $\alpha^{l}_{ij}$ is the attention coefficient of edge $e_{ij}$ of $l_{th}$ layer, $\lambda$ is attention coefficient, and $\omega^{l} $ is the corresponding input linear transformation’s weight matrix in $l_{th}$ layer. To increase the capacity of attention mechanism, GAT utilizes multi-head attentions for feature aggregation. In a multi-head attention block, each head works independently and all the outputs are concatenated to form a new feature representation for the next layer. Evidence suggests that the GAT enables the model to leverage sparse matrix operations, reducing storage complexity to linear in the number of nodes and edges.\\

\subsection{Biomedical knowledge graph}
Knowledge graph (KG) [20] plays an important role for knowledge-driven applications, which provides structural relationships among multiple entities and unstructured semantic relationships related to each entity. The knowledge graph is widely used in the biomedical field as biomedical knowledge graphs [21][22]. These graphs represent biomedical concepts and relationships in the form of nodes and edges [23]. The KG is comprised of entity-relation-entity triples and can be denoted by $KG=(H, R, T)$, where $H$ and $T$ denote different entities, and $R$ denotes the relationship types between $H$ and $T$. The biomedical knowledge graph (BKG) is utilized to obtain biomedical features, which is constructed from various biomedical data. In this study, we will adapt BKG to extract drug feature representations.\\

\section{Method}
\subsection{Problem Formulation}
In DDI prediction tasks, given a set of drugs $D=\{d_{1}, d_{2}, \dots d_{i}, \dots d_{j}, \dots d_{n}\}(1\leq i,j\leq n)$,  $n$ is the number of drugs. The main task of this study is to construct biomedical knowledge graph (BKG) from the given drugs, and find whether two drugs $d_{i}$ and $d_{j}$ have potential interactions. The BKG is defined as $\zeta_{DDI}=\{(d_{i},r_{ij},d_{j})|d_{i},d_{j}\in D;r_{ij}\in R\}$, where $r_{ij}$ is the biomedical interaction type among the entities $R$. $(d_{i},r_{ij},d_{j})$ describes a drug-drug interaction $r_{ij}$ between drug $d_{i}$ and $d_{j}$. The drug-drug interaction prediction aims to output the latent effect of the given drugs. More specifically, it aims to learn a prediction function $f_{ij}=F(d_{i},d_{j}|D,\zeta_{DDI},R)$ from drug pair $(d_{i},d_{j})$ and the interaction $r_{ij}\in R$. In this study, the function $f_{ij}\in \{0, 1\}$ predicts whether there is an interaction between a pair of drugs $(d_{i},d_{j})$ or not.\\

\subsection{ Overall framework of RGDA-DDI}

The proposed RGDA-DDI is a combination of a series of residual GAT blocks and a global residual concatenation for drug feature representation, which learns drug substructures from multi-scale feature aggregation and predicts DDIs via dual-attention blocks of all substructures of two drugs. The overall framework of RGDA-DDI is illustrated in Figure 1(a). Taking two single drug graphs as input, the residual-GAT module extracts both sub-structure and global-structure representations, respectively. Subsequently, the global-structure drug representations and each generated substructure drug representation embeddings propagate as input into the dual-attention layers to learn valuable features for positive predictions. Finally, the DDI prediction block is adopted to predict DDIs.\\
\begin{figure}[h!] 
  \centering 
  \includegraphics[width=1\textwidth, height=0.4\textheight]{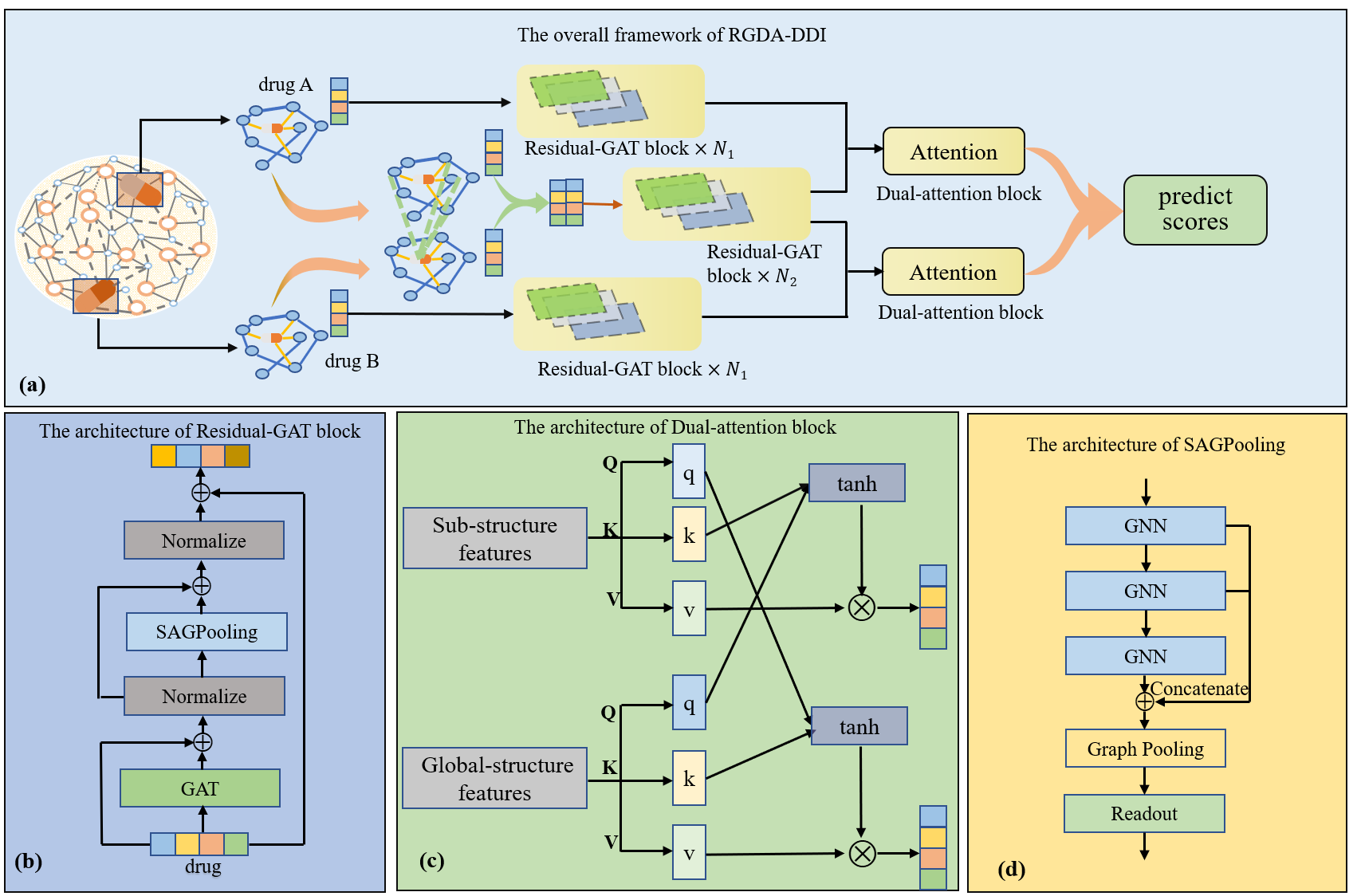} 
  \caption{A general figure of RGDA-DDI. (a)The overall framework of RGDA-DDI. (b)Residual-GAT block. (c)Dual-attention block. (d)The architecture of SAGPooling.} 
  \label{fig:example} 
\end{figure}\\
\subsubsection{BKG construction and embedding}
Inspired from BioDKG [24] and KGNN [25], we construct biomedical knowledge graphs by collecting raw data (e.g., DrugBank, KEGG, PharmGKB, or OFFSIDES) from corresponding portals and converted them to RDF graph using Bio2RDF tool [26]. Then, the RDF graph is uploaded to the RDF triple store, and extracts the selected triples by executing the federated SPARQL queries based on the billion triples benchmark [27]. Consequently, the extracted triples in the form of (entity, relation, entity) are constructed. Note that, here we created two data sources from the DrugBank dataset: (i) the parsed DDI matrix that contains the drug-drug pairs; and (ii) the constructed BKGs. Subsequently, to facilitate machine learning for fixed-length vector processes, a KG embedding procedure is necessary to be performed by encoding the BKGs into vectors. In this study, we employ the OpenKE tool [28] to perform KG embedding tasks according to previous methods [29], such as representing entities and relations, defining a scoring function, and learning entity and relation representation. Conceivably, these triples represent the neighborhood of a node as well as the kind of relations between itself and its neighboring nodes.\\
\subsubsection{Residual-GAT block for feature learning}
From Figure 1(a), a series of residual-GAT blocks have receptive fields at different scales for feature learning, using two drug graphs for sub-structure feature learning and a bipartite graph for global-structure feature learning as input. As illustrated in Figure 1(b), the residual-GAT block consists of a GAT encoder and a SAGPooling [30] layer which is concatenated by residual connection. For feature extraction, the GAT is employed to capture the sub-structure interaction representations. More specifically, the attentional weights are shared to each other along edges in the drug graph, each atom then aggregates all the information sent to the SAGPooling layer after normalization. \\

The self-attention graph pooling SAGPooling layer is utilized as a readout function of residual-GAT block. As profiled in Figure 1 (d), the SAGPooling uses self-attention and graph convolution to consider both node features and graph topology, which contains three graph convolutional layers where the outputs of each layer are concatenated. Finally, a readout layer is introduced to summarize the output features as shown in Eq. (2).\\
\begin{equation}
r=\frac{1}{N}\sum^{N}_{i=1} x_{i} \oplus M_{i=1}^{n} x_{i})
\end{equation}
where $N$ denotes the number of nodes, $M$ denotes the maximum value of the feature vectors from the $i_{th}$ node $x_{i}$, and $\oplus$ is the concatenation operation.\\
\subsubsection{Dual-attention block for feature aggregation}
In this study, we exploited a dual-attention block for feature aggregation using attention machines as stated in Figure 1(c) [31]. The attention weights can be calculated by three adaptive weights as shown in the following function (3).\\
\begin{equation}
\omega=softmax(\frac{Q\cdot K^{T}}{\sqrt{D_{K}}}\cdot V)
\end{equation}
where $Q$, $K$, $V$ denote the matrices that represent queries $Q\in R^{N\times D_{K}}$, keys $K\in R^{M\times D_{V}}$ and values $V\in R^{M\times D_{V}}$, $N$ and $M$ denote the lengths of $Q$ and $K$ or $V$, $D$ denotes the dimensions of the given matrix. The output attention weights $\Tilde{\omega}$ of the dual-attention block satisfies the formula shown below.\\
\begin{equation}
\Tilde{\omega}=(\omega_{Q}, \omega_{K}, \omega_{V})=\omega(Q, K, V)
\end{equation}
tanh function \\
\begin{equation}
\Tilde{\omega}=tanh\frac{Q_{1}\cdot K^{T}_{2}}{\sqrt{D_{K}}}\cdot V_{1}
\end{equation}
where $(\omega_Q,\omega_K,\omega_V)$ represents three adaptive attention weights.\\

\subsubsection{DDI prediction block}
At this stage, the embeddings of all the nodes in the heterogeneous network are obtained and these aggregations will be fed into the DDI prediction block to predict DDIs. For binary DDI prediction tasks, the interaction score is calculated by the sigmoid function as shown in Eq. (6).\\
\begin{equation}
f^{b}_{ij}=sigmoid(v_{i}||v_{j})
\end{equation}
where $f^{b}_{ij}\in \{0,1\}$, $v_i, v_j$ denotes the representation of drug $d_i$ and $d_j$, respectively. \\

\section{Experiment}
\subsection{Datasets}
To verify the effectiveness and progressiveness of the proposed RGDA-DDI model, a series of experiments were performed on the two DDI datasets respectively. We utilize the KEGG and DrugBank datasets to construct BKGs. The KEGG dataset contains 1,925 approved drugs and 56,983 interactions, and the DrugBank dataset contains 2,578 approved drugs and 612,388 interactions. To make the model calculation more convenient, we randomly generated the same number of negative samples as that of positives, where the negative sample is defined as the drug pairs that have not appeared in the positive ones. We then employed the OpenKE tool for BKG construction. The datasets used in this study are summarized in Table 1.\\
\begin{table}
\caption{Details of adopted datasets in this study}
\begin{center}
\renewcommand{\arraystretch}{1.3}
\begin{tabular}{c@{\quad}c@{\quad}c}
\hline
\multicolumn{1}{l}{\rule{0pt}{12pt}} & \multicolumn{1}{l}{DrugBank} & \multicolumn{1}{l}{KEGG-drug}\\[2pt]
\hline\rule{0pt}{12pt}
Drugs  & 2578  &  1925\\
Interactions  & 612,388  &  56,983\\ 
Entities  & 2,129,712  &  129,910\\ 
Relation types  & 72  & 167\\ 
KG triples  & 7,852,852  & 362,870\\
\hline
\end{tabular}
\end{center}
\end{table}

\subsection{Baselines}
There are several state-of-the-art baselines used, such as the LINE model that embeds graphs in network-based methods, GCN and GAT based DDI prediction methods, Deepwalk [32] and Deep-DDI [8]. (which are the CNN-based DDI prediction models) In addition, we reconstructed some recently published DDI prediction models such as LaGAT [15], KGNN [25], and MUFFIN [33] for comparison. LaGAT uses the embedding representation of one of the drugs as a query vector to calculate the attention weights, and selects the appropriate topological neighbor nodes to obtain the semantic information of the other drug. KGNN extracts drug features through GNN and external KG, the neighborhood information of each node is sampled and aggregated from the local receiver of each node for DDI prediction. MUFFIN is a multi-scale feature fusion deep learning model; it learns drug representations based on both the drug-self structure and the KG with rich bio-medical information. In addition, to make relatively fair discussion and comparison, we keep the rest of the parameters consistent with our proposed method.\\

\subsection{Evaluation metrics}
To evaluate the performance of the proposed RGDA-DDI model and the baselines comprehensively, a series of evaluation metrics such as accuracy (ACC), area under the receiver operating characteristic curve (AUC), area under the precision-recall curve (AUPR), and F-score are used in this study. The ACC is used to measure the proportion of correctly predicted samples to the total numbers. AUC considers the performance of classifiers for both positive and negative predictions, and it can still make reasonable evaluations even in the case of an imbalanced dataset. AUPR is used to evaluate the balance between precision and recall under different prediction thresholds. F-score takes both precision and recall into consideration, as the score improves only when both metrics are high. Furthermore, for fairness, the stratified 5-fold cross-validation strategy is implemented in the training process to avoid data biases.\\

\subsection{Implementation details}
This experiment was implemented on CentOS 7.5 Linux operating system. We employed Python 3.6 language with TensorFlow 2.2. framework. The number of residual-GAT blocks is set to 10 (N1=N2=10). Each block has a representation layer with the shared weights for each drug and both two drugs. All layers adopt a GAT mechanism equipped with two attention heads for message passing. The RGDA-DDI first encodes drugs into 1 by 64 vectors. The combination of the local feature learning for individual drug information and the global feature learning for drug pairs generates a 2 by 64 graph vector as the model input. The block layers for individual drugs are set to 10, while the block layers for drug pairs are set to 8. The binary cross-entropy (BCE) loss function being used is demonstrated in Eq.(6).\\
\begin{equation}
Loss_{b}=\sum_{(d_i,d_j)\in R}(\Tilde{f_{ij}}log(f_{ij}))+(1-\Tilde{f_{ij}})log(1-\Tilde{f_{ij}})
\end{equation}
where $\Tilde{f_{ij}}\in \{0,1\}$ denotes the true DDI value, and $f_{ij}\in \{0,1\}$ is the predicted value. The 5-fold cross-validation is performed in this study, where we randomly divide the dataset into 5 folds, 4 of which are training sets, and the remaining one is equally divided into validation set and test set. We used the Adam SGD optimizer and trained the models with 100 epochs on DrugBank and 120 epochs on KEGG-drug. The learning rate was set to 0.012, and the batch size was set to 1024.\\

\subsection{Results}
\subsubsection{Result analysis}
To compare the performance of RGDA-DDI with the baselines fairly and comprehensively, we performed the experiments across 5 runs. The experimental results are detailed in Table 2. With the DrugBank dataset, the AUC, ACC,	F1-scores of the proposed RGDA-DDI model are the best compared to all baselines, except that AUPR is slightly behind LAGAT. The overall advantage of RGDA-DDI is still obvious. On the other hand, on the KEGG-drug dataset, RGDA-DDI model achieves the best performance compared to all baselines on these metrics. The RGDA-DDI improves by about \% on AUC compared to Deepwalk, and about 50\% on F1-Score and 15\% on ACC compared to the LINE model. These results demonstrate the effectiveness and progressiveness of the proposed model. One possible reason is that the residual-GAT module extracts multi-scale drug representations and the dual-attention layers to learn the valuable features for positive predictions. It can be drawn that the performance of the proposed model is impacted by the scale of datasets. When trained on the DrugBank dataset, the RGDA-DDI performs on par with LAGAT. However, on the smaller KEGG-drug dataset, our proposed RGDA-DDI has achieved significant advantages in all metrics, which proves its powerful ability in feature extraction of small sample data.\\

\begin{table}[h] 
\caption{Experimental results, bold: best results}
\begin{center}
\renewcommand{\arraystretch}{1.3}
\begin{tabular}{c@{\quad}c@{\quad}c@{\quad}c@{\quad}c@{\quad}c}
\hline
\multicolumn{1}{c}{Dataset}{\rule{0pt}{12pt}} & \multicolumn{1}{c}{Model} & \multicolumn{1}{c}{AUC} & \multicolumn{1}{c}{ACC} & \multicolumn{1}{c}{F1-score} & \multicolumn{1}{c}{AUPR} \\[2pt]
\hline\rule{0pt}{12pt}
         & LINE  &  0.8145 & 0.7657 & 0.3026 & 0.4235 \\
         & GCN  &  0.7962 & 0.7541 & 0.7614 & 0.7963 \\
         & GAT  &  0.8521 & 0.7641 & 0.7762 & 0.8204 \\
         & Deepwalk  &  0.7240 & 0.6850 & 0.2477 & 0.2951 \\
DrugBank & Deep-DDI  &  0.8045 & 0.7602 & 0.7871 & 0.8262 \\ 
         & LAGAT  &  0.9262 & 0.8592 & 0.7899 & \textbf{0.8349} \\
         & KGNN  &  0.9203 & 0.9089 & 0.7399 & 0.7195 \\
         & MUFFIN  &  0.9156 & 0.8842 & 0.7749 & 0.7601 \\
         & \textbf{RGDA-DDI} &  \textbf{0.9341} & \textbf{0.9105} & \textbf{0.7991} & 0.8091 \\
\hline\rule{0pt}{12pt}
 & LINE  &  0.9264 & 0.8655 & 0.8695 & 0.8968 \\
          & GCN  &  0.9652 & 0.9043 & 0.9182 & 0.9414 \\
          & GAT  &  0.9742 & 0.9288 & 0.9297 & 0.9667 \\
          & Deepwalk  &  0.8850 & 0.8130 & 0.8170 & 0.7970 \\
KEGG-drug & Deep-DDI  &  0.9748 & 0.9166 & 0.9167 & 0.9161 \\ 
          & LAGAT  &  0.9796 & 0.9490 & 0.9296 & 0.9795 \\
          & KGNN  &  0.9518 & 0.8950 & 0.9032 & 0.9533 \\
          & MUFFIN  &  0.9797 & 0.9261 & 0.9265 & 0.9175 \\
          & \textbf{RGDA-DDI} &  \textbf{0.9864} & \textbf{0.9506} & \textbf{0.9310} & \textbf{0.9822} \\
\hline
\end{tabular}
\end{center}
\end{table}

\subsubsection{Ablation experiment}
To verify each component of the proposed model, we conducted the ablation experiment in this work. The embodiments are detailed in Table 3. We first deleted the attention block and simplified the concatenation of the drug features, which is marked as Sub-model A. Furthermore, we canceled the residual connections in the residual-GAT block and marked it as Sub-model B. Finally, we deleted the SAGPooling block and marked it as Sub-model C. The Sub-models are performed on the two datasets respectively. From the experimental results, it can be found that Sub-model A has the worst performance, proving that the attention blocks are very effective and crucial to the model’s performance. In addition, the other components also impact the performance as the result shows, but not as significant as the attention block. All of the sub-models fully adopt the biochemical features that ensure the scalability to the proposed model, but only when integrating all components, the performance is the best.\\
\begin{table}[H]
\caption{Results of ablation test on RGDA-DDI, bold: best results}
\begin{center}
\renewcommand{\arraystretch}{1.3}
\begin{tabular}{c@{\quad}c@{\quad}c@{\quad}c@{\quad}c@{\quad}c}
\hline
\multicolumn{1}{c}{Dataset}{\rule{0pt}{12pt}} & \multicolumn{1}{c}{Sub-Model} & \multicolumn{1}{c}{AUC} & \multicolumn{1}{c}{ACC} & \multicolumn{1}{c}{F1-score} & \multicolumn{1}{c}{AUPR} \\[2pt]
\hline\rule{0pt}{12pt}
DrugBank & Sub-model A  &  0.7283 & 0.6841 & 0.5642 & 0.7543 \\
         & Sub-model B  &  0.8085 & 0.7956 & 0.7448 & 0.7426 \\
         & Sub-model C  &  0.8892 & 0.8999 & 0.7665 & 0.7805 \\
         & \textbf{Proposed model}  &  \textbf{0.9341} & \textbf{0.9105} & \textbf{0.7991} & \textbf{0.8091} \\
\hline\rule{0pt}{12pt}
KEGG-drug & Sub-model A  &  0.9419 & 0.8985 & 0.8982 & 0.9547 \\
          & Sub-model B  &  0.9426 & 0.9007 & 0.9021 & 0.9435 \\
          & Sub-model C  &  0.9509 & 0.9071 & 0.9077 & 0.9528 \\
          & \textbf{Proposed model}  &  \textbf{0.9864} & \textbf{0.9506} & \textbf{0.9310} & \textbf{0.9822} \\
\hline
\end{tabular}
\end{center}
\end{table}

\subsubsection{Discussion}
In this section, we examine the influence of the number of different residual-GAT blocks on the performance of the proposed method. The RGDA-DDI model contains two types of residual-GAT blocks, N1 represents the number of single drug residual-GAT blocks, and N2 represents the number of DDP residual-GAT blocks. We performed experiments on the KEGG-drug dataset and set the number of residual-GAT blocks as 2, 5, 8, 10, 12, respectively. The AUC results of different N1and N2 values are detailed in Table 4. From the experimental results we can see that the model performs best when both N1 and N2 are 10. This is because a too small N1 or N2 does not have enough capacity to incorporate drug information, while larger values are prone to be misled by noises.\\
\begin{table}
\caption{The AUC results of different N1 and N2 values, bold: best results}
\begin{center}
\renewcommand{\arraystretch}{1.3}
\begin{tabular}{c@{\quad}c@{\quad}c@{\quad}c@{\quad}c@{\quad}c}
\hline
\multicolumn{1}{c}{\backslashbox{N2}{N1}}{\rule{0pt}{12pt}} & \multicolumn{1}{c}{2} & \multicolumn{1}{c}{5} & \multicolumn{1}{c}{8} & \multicolumn{1}{c}{10} & \multicolumn{1}{c}{12} \\[2pt]
\hline\rule{0pt}{12pt}
2   & 0.9541  & 0.9579 & 0.9672 & 0.9719 & 0.9698 \\
5   & 0.9643  & 0.9621 & 0.9701 & 0.9621 & 0.9642 \\
8   & 0.9688  & 0.9621 & 0.9676 & 0.9754 & 0.9711 \\
10  & 0.9712  & 0.9721 & 0.9742 & \textbf{0.9764} & 0.9701 \\
12  & 0.9643  & 0.9762 & 0.9718 & 0.9749 & 0.9686 \\
\hline
\end{tabular}
\end{center}
\end{table}

\vspace{-25 pt}
\section{Conclution}
In this study, we proposed a novel drug-drug interaction prediction framework named RGDA-DDI. RGDA-DDI can adaptively learn and integrate multi-scale drug feature representation, and a dual-attention block is designed to grasp the key features that contribute to the positive predictions. The proposed novel architecture achieves better performance than the existing state-of-the-art baselines on a series of evaluation metrics. In addition, the RGDA-DDI has achieved significant performance improvement in a more challenging inductive scenario, with an improvement of about 50\% on F1-Score and 15\% on ACC, compared to the LINE model. By performing multi-scale drug-drug interaction features and dual-attention for feature fusions, we have demonstrated the power of integrating joint drug-drug information during DDI prediction tasks. In a nutshell, the novel RGDA-DDI provided a new methodology in the field of drug development. Future research could further improve the molecular feature representation capability of drugs to prepare for the discovery of new drugs.\\
\paragraph{Conflict of interest.}
The authors declared no conflict of interest.

\paragraph{Author contribution.}
Changjian Zhou: Conceptualization, Methodology, Coding, Writing original draft. Xin Zhang: Data curation, Methodology, Coding. Jiafeng Li: Advisory, Manuscript Review, Revisions.  Jia Song and Wensheng Xiang: Conceptualization, Methodology, Supervision, and Funding.

\paragraph{Notes and Comments.}
The authors would like to thank colleagues and the anonymous reviewers who have provided valuable feedback to help improve this paper. The data that support the findings of this study are available from \textit{https://github.com/HPC-NEAU/RGDA-DDI}. This work was supported by the National Key Research and Development Program of China \footnote{No. 2023YFD1700700.}.\\

%
%

%
\end{document}